\documentclass{svjour3}                      
\smartqed  
\usepackage{graphicx}
\usepackage{epstopdf}
 \journalname{Accepted in Journal of Optics (2013) of}
\begin{document}

\title{Artificial color perception using microwaves 
}


\author{Debesh Choudhury         \and
        H. John Caulfield 
}


\institute{Debesh Choudhury \at
              Department of Electronics and Communication Engineering\\
              JIS College of Engineering,
              Block A, Phase III, Kalyani, Nadia, 
              Pin 741235,\\ West Bengal, India. \\
              \email{debesh@iitbombay.org}
           \and
           H. John Caulfield$^{\dagger}$ \at
              Alabama A\&M University Research Institute\\
              P.O. Box 313, Normal, AL 35762, USA\\
              \emph{Present address:} \at 
              Debesh Choudhury\\
              Department of Electronics and Communication Engineering\\
              Neotia Institute of Technology, Management and Science\\
              PO - Amira, D. H. Road, South 24 Parganas, 
              Pin 743368, West Bengal, India.
}

\date{}


\maketitle

 \begin{abstract}
 We report the feasibility of artificial color perception under 
microwave illumination using a standard microwave source and an antenna. 
We have sensed transmitted microwave power through color objects and 
have distinguished the colors by analyzing the sensed transmitted power. 
Experiments are carried out using a Gunn diode as the microwave source, 
some colored liquids as the objects and a microwave diode as the detector. 
Results are presented which open up an unusual but new way of perceiving 
colors using microwaves.
 \vglue 12pt
 \keywords{Color sensing \and Artificial color \and Microwave imaging}
 \end{abstract}

\section{Motivation}   \label{motive}
 The perception of color is not unique in the nature. Color vision 
varies from animal to animal. It even changes from person to person. As 
for example, a color blind person may perceive a color differently in 
some scene than a `normal' human being. In other words, color perception 
depends on specific color filtering abilities in the eyes and also upon 
the brain processes that create the color in their percepts of the 
scene. Humans and other animals\footnote{So far as we know, only the 
very remarkable mantis shrimp has that capability.} cannot change or 
adjust these color filters provided by their cone cells. On the other 
hand, the color filters in an artificial eye, e.g., of a camera, may be 
customized according to the needs. Thus, color sensing of a camera can 
be customized to yield a color perception which may be called artificial 
color~\cite{hjc_nc03,hjc_ei04}. Here the word `artificial' simply means 
something humans rather than nature has chosen.

In fact, images captured by ordinary color cameras may be processed on 
the computer to achieve results of artificial 
colors~\cite{hjc_ivc05,hjc_ist05}, which can improve applications in 
binary logic and biometric recognition~\cite{hjc_is04,hjc_prl05}. A 
question arises, whether one can perceive colors beyond the visible band 
of wavelengths. In what follows, we try to address the problem of 
artificial color perception using microwave 
radiation~\cite{dc_hjc_icontop11}.

\section{Introduction}   \label{intro}

Color is generally referred to as corresponding to the different visible 
wavelength bands. But that view is plainly approximate and not very 
useful. Color is not something in the world awaiting detection. Rather 
it is the result of brain processing to provide spectral discrimination. 
A visual percept of a scene may contain objects at various 3D locations 
in the scene but it also contains spectral discriminants calculated by 
the brain and attributed to objects in the percept. Those discriminants 
are what we call color. Color perception comes to reality by the 
response of the eye to these visible radiations. In Sun light, Newton 
found seven colors but could not find ultra-violet, infrared and 
microwaves, because only visible wave bands can be seen by us. Thus, the 
concept of color has never been associated with ultra-violet, infrared 
or microwave radiations. But color and spectrum are not the same kinds 
of things. Spectra come from the outside world and colors come from the 
brains of animals that sense some of that spectral information.

Microwave sources radiate a wide band of frequencies of 1 GHz to 100 GHz 
and the wavelength varies from 30 cm to 0.33 cm. On the contrary, 
optical radiation fall in the frequency range of 670 THz to 450 THz 
corresponding to wavelengths of 0.4 $\mu$m to 0.7 $\mu$m. Though both 
microwaves and optical waves are electromagnetic in nature, the 
different wavelengths produce different imaging characteristics. For 
example microwave radiation is capable of penetrating objects that are 
opaque to visible light. Microwaves are very dominant in communication 
systems where a variety of antennas can receive microwave 
signals~\cite{mw_book84}. Microwave radiations have a very popular 
application in our households as microwave cooking 
ovens~\cite{mw_paper_ieee84}. Microwave imaging is extensively used in 
medical diagnostics, building research and military 
applications~\cite{mwi_book09}. However, color perception in the 
microwave bands has not yet been found in the literature so far.

We explore artificial color perception in the microwave bands utilizing 
standard microwave antennas. A microwave antenna is proposed to sense 
the wave fields transmitted through an object having different visible 
colors into it. The microwave source may be a variable frequency 
source, such as a Klystron or a Gunn diode. Microwaves are affected by 
the dielectric properties of the matter, so transmission of microwave 
radiation through otherwise visual colored objects will show variations 
in transmitted power due to the variation of dielectric values in the 
object. Some intuitive experimentation suggests that microwaves may 
offer an unusual forte for sensing the otherwise visual color percepts 
into artificial microwave color percepts. To the best of our knowledge, 
this is the first effort of exploring color perception under microwave 
radiation.

\section{Intuitive Experiments}

There are a variety of microwave antennas available, such as horn, 
microstrip antennas etc. [8]. Among them horn antennas are commonly 
available. We propose to use a common horn antenna and a microwave diode 
detector to sense the transmitted microwaves through an object. We try 
some simple experiments using a Gunn diode as source.  Ideally, we wish 
to capture the transmitted microwave radiation by scanning the detector 
antenna on a
 \begin{figure}[h]
 \centering
  \includegraphics[width=3.5in]{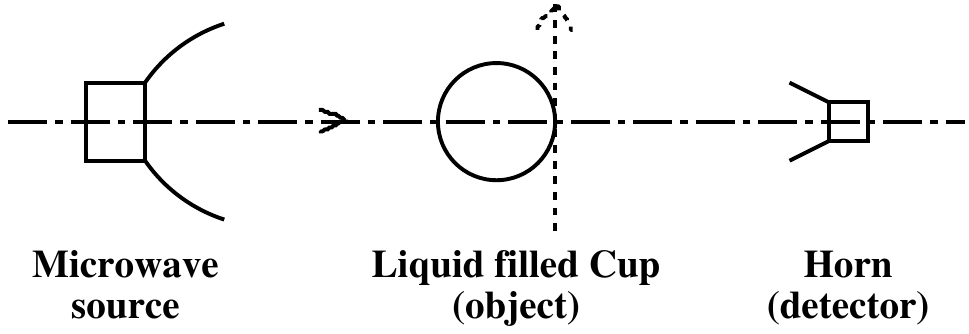}
 \caption{The schematics of the experimental set-up. The dotted arrow 
shows the direction along which the object is moved.
 \label{fig:1}
         }
 \end{figure}
 straight line orthogonal to the propagation axis. That means we are 
interested in capturing the transmitted radiation in one dimension only. 
In fact, we have moved the object instead of the antenna, because the 
antenna is fixed on the bench. This does not affect the sensing since we 
are interested to see how much amplitude of microwaves is transmitted by 
the object.

The schematic figure of the experimental setup is shown in 
Fig.\ref{fig:1}. The Gunn diode source is made to radiate at a frequency 
of approximately 10 GHz which is equivalent to a wavelength of 
approximately 3 cm. A small polythene cup filled with a liquid is placed 
as an object in front of the Gunn diode. The microwave radiation 
transmitted through the object is detected with a horn antenna and a 
diode detector. The detector is connected to a
 \begin{figure}[h]
 \centering
  \includegraphics[width=3.95in]{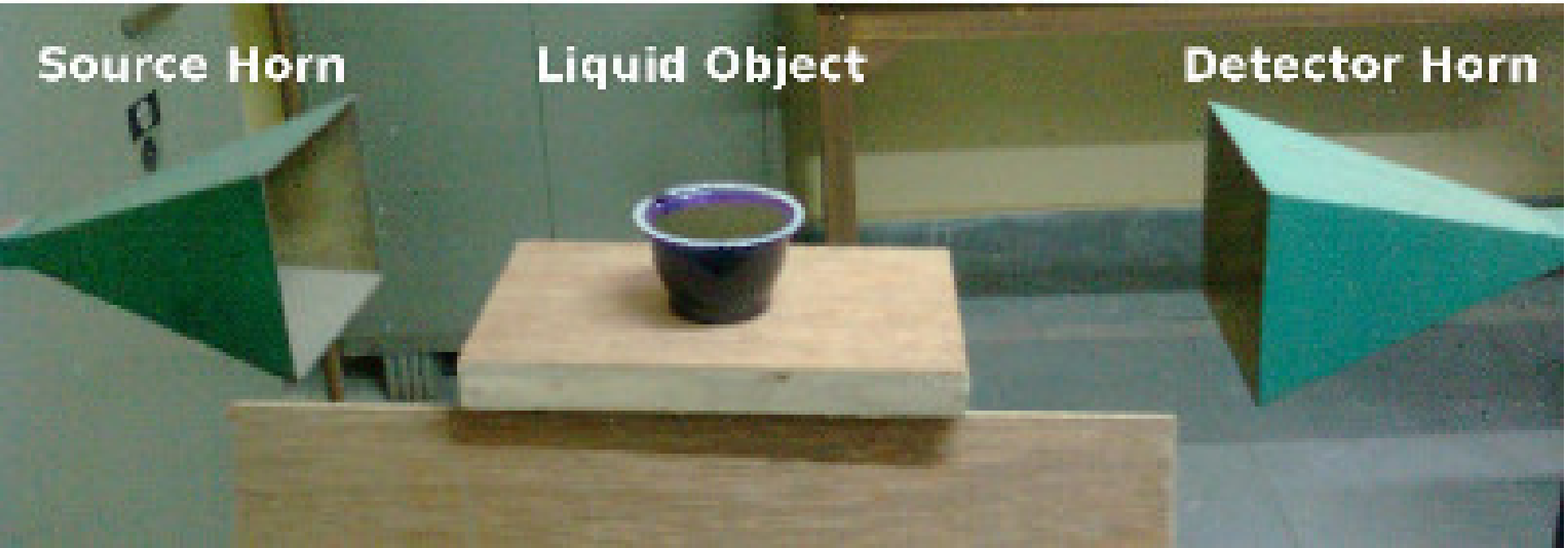}
 \caption{Photograph of the experimental arrangement.
            \label{fig:2}  
          }
 \end{figure}
 microwave power meter (not shown in Fig.1). The actual photograph of 
the experimental arrangement is shown in Fig.\ref{fig:2}. The polythene 
cup is filled with plain water and is moved orthogonal to the axis of 
propagation and subsequent data are recorded.  Now, the experiment is 
repeated after filling the cup with milk tea. The same experiment is 
repeated after
 \begin{figure}[h]
 \centering
  \includegraphics[width=0.9in]{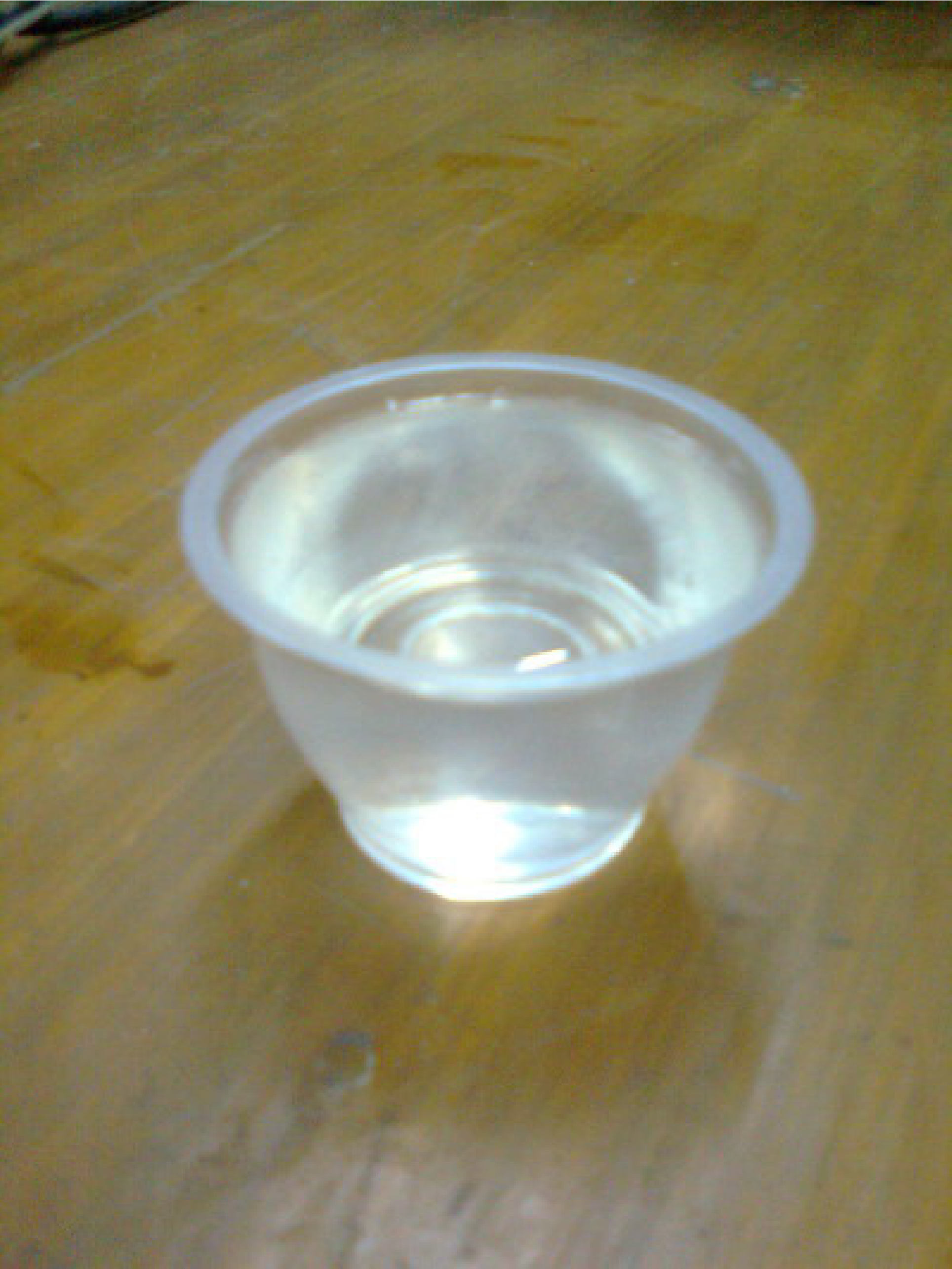}
  \hskip 0.5in
  \includegraphics[width=0.9in]{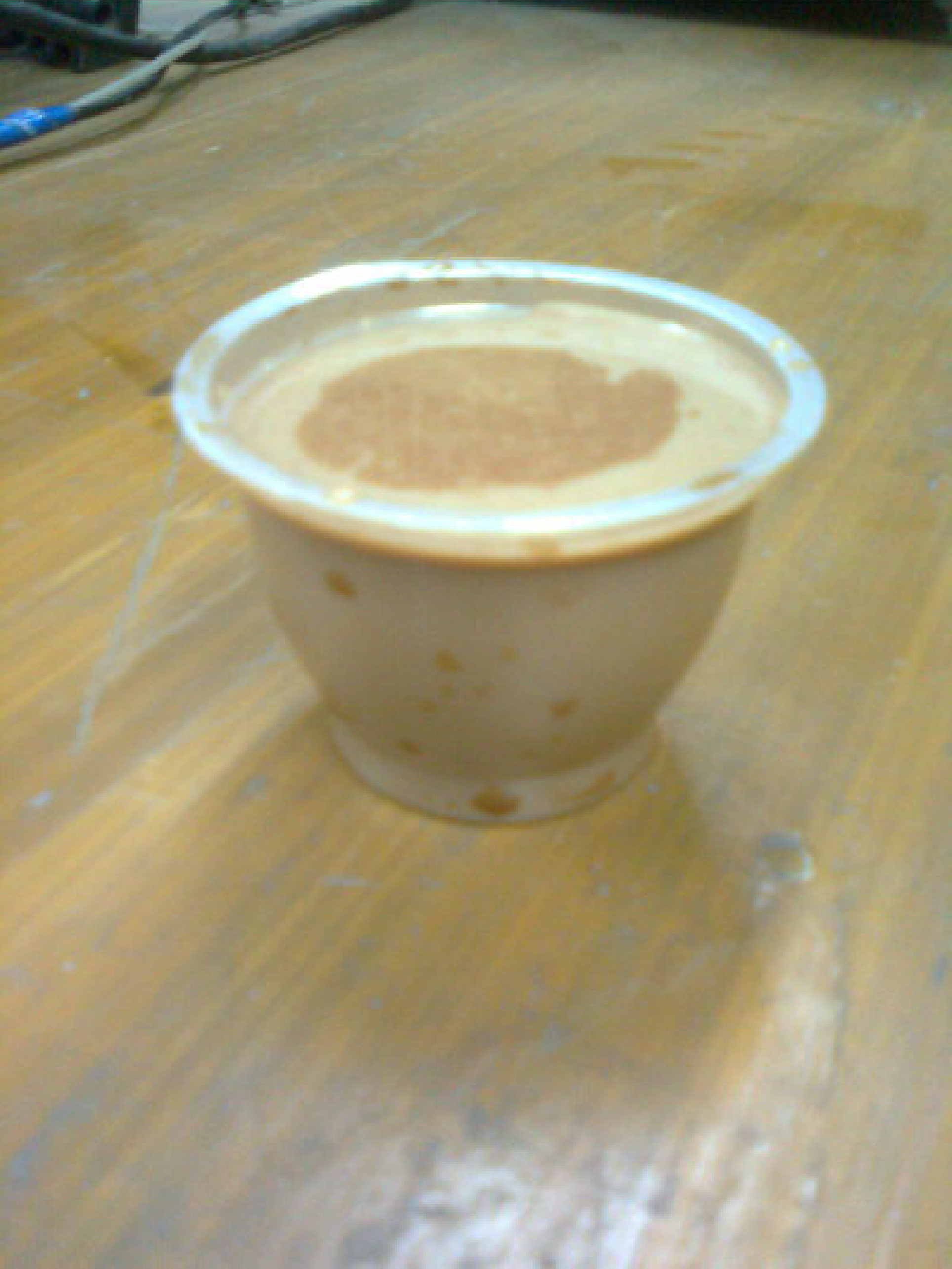}
  \hskip 0.5in
  \includegraphics[width=0.9in]{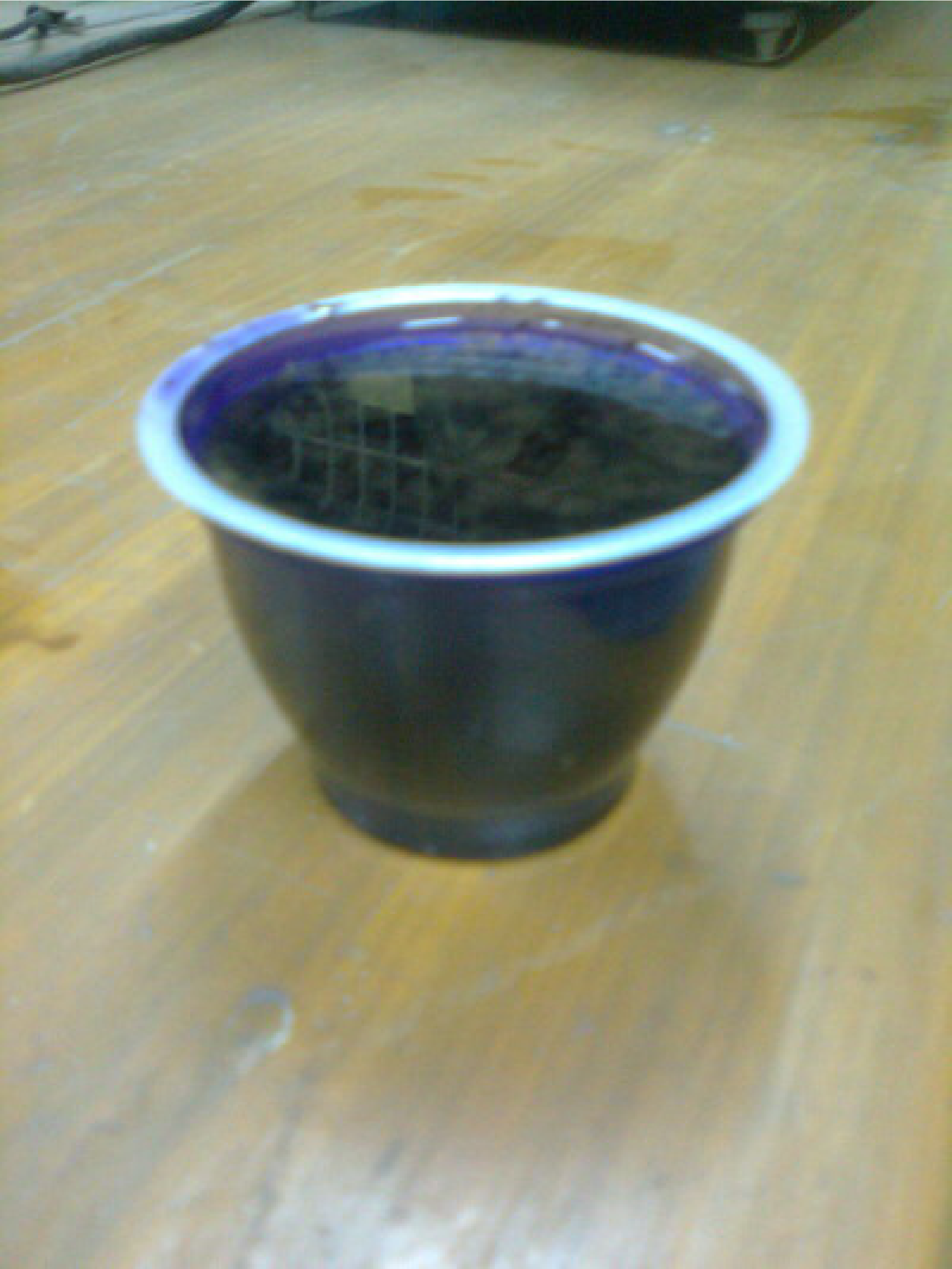} 
  \\[10pt] {\rm (a)}  \hskip 1.2in {\rm (b)} \hskip 1.2in {\rm (c)}
 \caption{Photographs of the objects (a)~cup filled with plain 
water, (b)~cup filled with milk tea, (c)~cup filled with water colored 
by blue marker pen ink.
  \label{fig:3}
         }
 \end{figure}
 filling the cup with colored water by blue ink of marker pen. The 
photographs of the objects (polythene cup filled with liquid) used in 
the experiments are shown in Fig.\ref{fig:3}. The photographs show the 
cup filled with plain water [Fig.3(a)], cup filled with milk tea 
[fig.3(b)] and the cup filled with water colored blue by marker pen ink 
[Fig.3(c)]. The experimental data plots for the cup filled with liquid 
is shown in Fig.4. The plot of the experimentally recorded transmitted 
microwave
 \begin{figure}[h]
 \centering
  \includegraphics[width=2.3in]{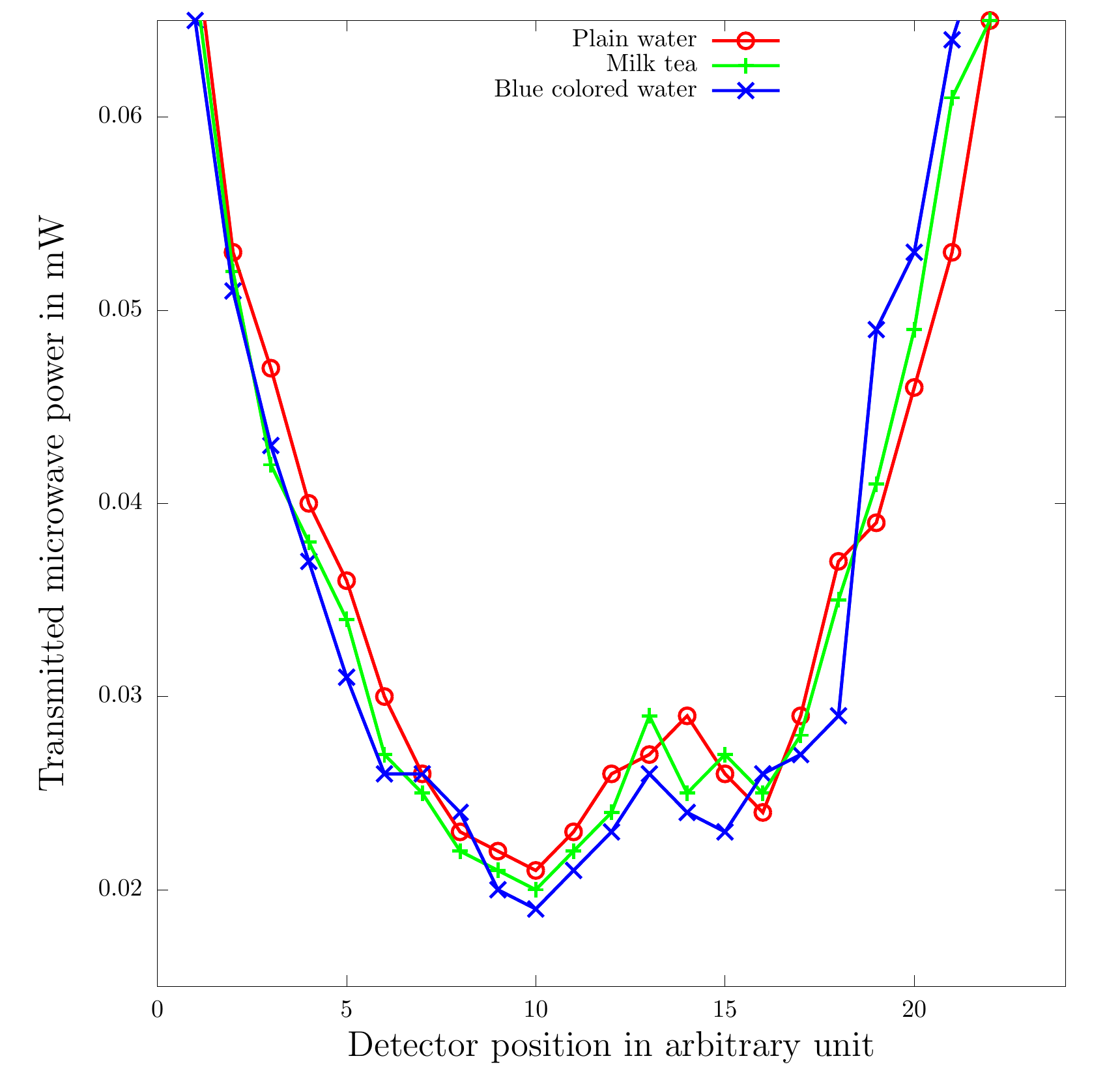}
  \hskip 0.1in
  \includegraphics[width=2.3in]{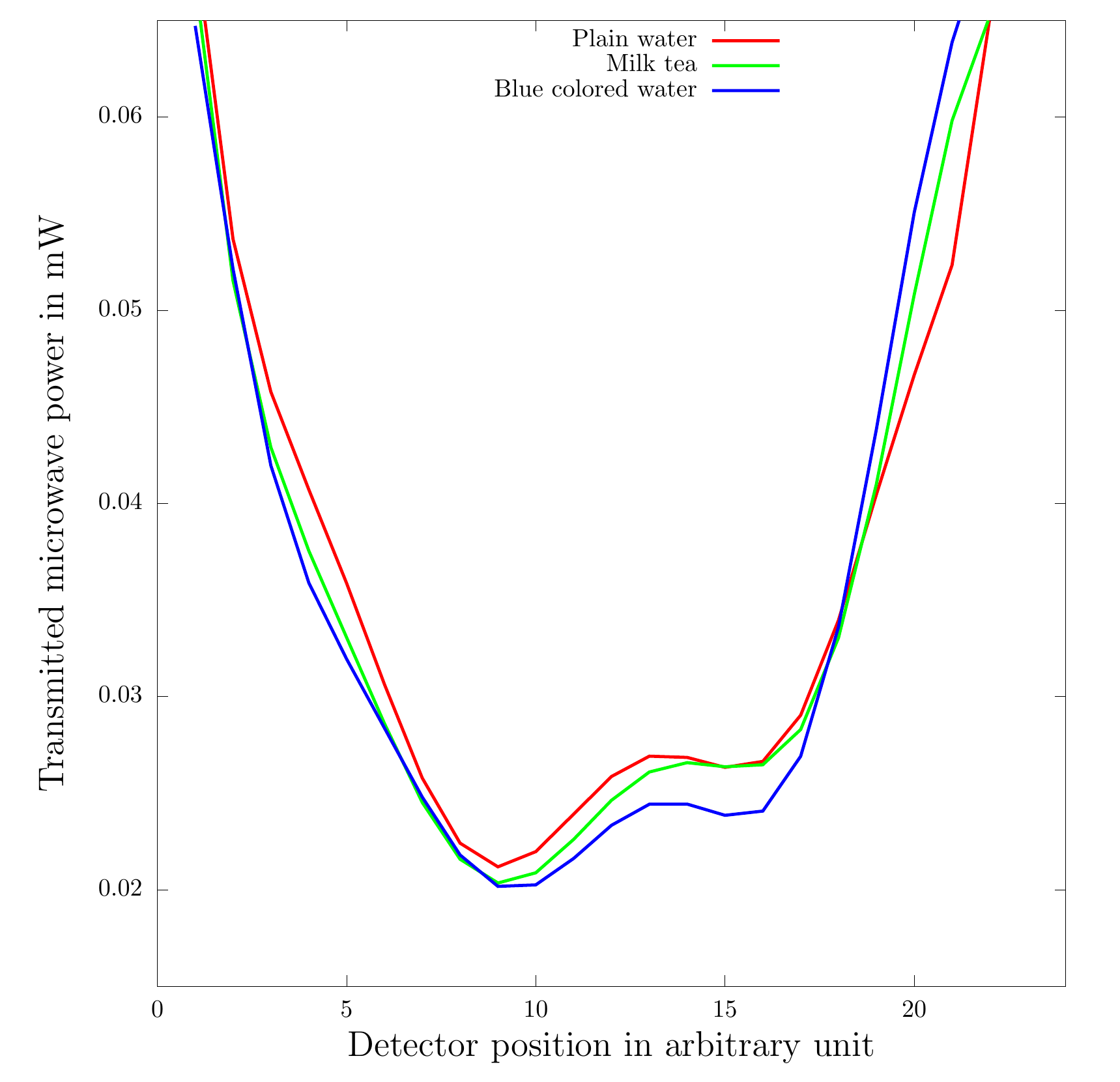}
  \\[10pt] {\rm (a)}  \hskip 2.2in {\rm (b)}
 \caption{Plots of transmitted microwave power (mW) versus detector (or 
object) position in arbitrary unit for (i)~plain water (red), (ii)~milk tea 
(green), (iii) blue colored water (blue). (a)~Plots with data points, 
(b)~Plots with curve fitting. 
 \label{fig:4}
        }
 \end{figure}
 power in mW versus the lateral position (in arbitrary unit) of the cup 
filled with plain water is shown in red color, that for the cup filled 
with milk tea is shown in green color and that for the cup filled with 
blue marker pen ink colored water is shown in blue color. Figure 4(a) 
shows the plotted data points, whereas Fig.4(b) is the same data plotted 
with polynomial curve fitting. It is discernible from Fig.4 that the 
microwave transmission due to plain water filled polythene cup and due 
to the polythene cup filled with milk tea and blue colored water are not 
same.

\section{Analysis of the Results}

The results (Fig.4) of the experiments with liquid as the transmissive 
object shows difference in the transmitted microwave power through plain 
water, milk tea and blue colored water. Since, microwave is affected by 
the dielectric properties, and the result is an effect of dielectric 
properties of the dissolved materials (milk, tea and color) in the 
water. The dielectric constant of water is very high ($\sim 73$), and adding 
milk tea and blue ink to it yields a finite changes in the dielectric 
values. The differences in the readings due to different colored water 
are due to the relative changes in dielectric values. The bending of the 
plots at the right hand bottom may be accounted for probable dielectric 
lensing action of the water filled cups on the microwaves. It may be 
noted that the same color on different objects with different dielectric 
properties may also result in ambiguities in the sensed colors.

We may call these unusual color percepts as microwave sensitive 
dielectric color, or simply dielectric color. These can be viewed as 
color detection sensitivities in microwave artificial color. As shown in 
Fig.4(a) and Fig.4(b), they are so similar as to be of little value, but 
they are distinctly different. The difference needs to be amplified and 
the similarity decreased. Those distinct differences in the three curves 
of Fig.4(a) and Fig.4(b) can be converted into three orthonormal signals 
using the Caulfield-Maloney filter~\cite{hjc_AO_69}, or Gram-Schmidt 
orthogonalization~\cite{gso_66}. These curves provide color specific 
sensitivities which are useful for looking at the three kinds of liquids 
in a scene. If the result is to be displayed to a human, an arbitrary 
mapping between the microwave sensitivity curves and the normal human 
three color sensitivity curves will enable that.

The results of our experiments prove the feasibility of using microwave 
radiation for discriminating visible color information of objects that 
can produce a sufficient change in its dielectric properties. The 
possibility of detecting color signals may be improved by using more than 
one microwave frequencies. The more elaborate experimentation need some 
dedicated sources, sensors and test benches which presently we don't 
have.

\section{Conclusion}

We have proposed color sensing and perception using microwaves. 
Empirical experimental studies show that artificial color in the 
microwave domain can be achieved very simply using readily available 
antennas. It can produce results different from conventional microwave 
images in that far more than simple intensity is displayed. The spectral 
`meaning' of the received radiation may be displayed as in any other 
color display, be it natural or artificial.

 \begin{acknowledgements}
 The authors are indebted to Mr. G. Das and Mr. G. Lohar for assistance 
in experiments. Logistic supports have also been received from Mr. S. 
Das and Mr. R. Sengupta.
 \end{acknowledgements}

\vspace{36pt} \noindent $^{\dagger}$ Professor H. John Caulfield is no 
more with us. His sole rests in peace in heaven since 31 January 2012.

\end{document}